# Rapid 2D $^{23}$Na MRI of the calf using a denoising convolutional neural network

Rebecca R. Baker[a,b], Vivek Muthurangu[b], Marilena Rega[c], Stephen B. Walsh[d], Jennifer A. Steeden[b]

[a] UCL Centre for Medical Imaging, University College London, London, UK
[b] UCL Centre for Translational Cardiovascular Imaging, University College London, London, UK
[c] Institute of Nuclear Medicine, University College Hospital, London, UK
[d] Department of Renal Medicine, University College London, London, UK

Author email addresses:
Rebecca R. Baker: r.baker.17@ucl.ac.uk
Vivek Muthurangu: v.muthurangu@ucl.ac.uk
Marilena Rega: marilena.rega@nhs.net
Stephen B. Walsh: stephen.walsh@ucl.ac.uk
Jennifer A. Steeden: jennifer.steeden@ucl.ac.uk

Corresponding author contact details:
Dr Rebecca Baker
UCL Centre for Translational Cardiovascular Imaging, University College London, Zayed Centre for Research, 20 Guilford St, London, WC1N 1DZ
r.baker.17@ucl.ac.uk

## Abstract

**Purpose:** $^{23}$Na MRI can be used to quantify in-vivo tissue sodium concentration (TSC), but the inherently low $^{23}$Na signal leads to long scan times and/or noisy or low-resolution images. Reconstruction algorithms such as compressed sensing (CS) have been proposed to mitigate low signal-to-noise ratio (SNR); although, these can result in unnatural images, suboptimal denoising and long processing times. Recently, machine learning has been increasingly used to denoise $^{1}$H MRI acquisitions; however, this approach typically requires large volumes of high-quality training data, which is not readily available for $^{23}$Na MRI. Here, we propose using $^{1}$H data to train a denoising convolutional neural network (CNN), which we subsequently demonstrate on prospective $^{23}$Na images of the calf.

**Methods:** 1893 $^{1}$H fat-saturated transverse slices of the knee from the open-source fastMRI dataset were used to train denoising CNNs for different levels of noise. Synthetic low SNR images were generated by adding gaussian noise to the high-quality $^{1}$H k-space data before reconstruction to create paired training data. For prospective testing, $^{23}$Na images of the calf were acquired in 10 healthy volunteers with a total of 150 averages over 10 minutes, which were used as a reference throughout the study. From this data, images with fewer averages were retrospectively reconstructed using a non-uniform fast Fourier transform (NUFFT) as well as CS, with the NUFFT images subsequently denoised using the trained CNN.

**Results:** CNNs were successfully applied to $^{23}$Na images reconstructed with 50, 40 and 30 averages. Estimated SNR was significantly higher in CNN-denoised images compared to NUFFT, CS and reference images. Quantitative edge sharpness was equivalent for all images. For subjective image quality ranking, CNN-denoised images ranked equally best with reference images and significantly better than NUFFT and CS images. Muscle and skin apparent TSC quantification from CNN-denoised images were equivalent to those from CS images, with less than 0.9 mM bias compared to reference values.

**Conclusion:** Denoising CNNs trained on $^{1}$H data can be successfully applied to $^{23}$Na images of the calf; thus, allowing scan time to be reduced from 10 minutes to 2 minutes with little impact on image quality or apparent TSC quantification accuracy.

# 1 Introduction

Sodium ($^{23}$Na) magnetic resonance imaging (MRI) can be used to quantify tissue sodium concentration (TSC), a potential biomarker for chronic kidney disease,[1] hypertension[2] and multiple sclerosis.[3] However, $^{23}$Na MRI is more technically challenging than proton ($^{1}$H) imaging due to the inherently lower signal-to-noise ratio (SNR). Low SNR can be compensated for by acquiring multiple signal averages or through three-dimensional (3D) acquisitions, but both significantly increase scan times. Unfortunately, long scan times limit clinical use of $^{23}$Na MRI and thus accelerated techniques are desirable.

Postprocessing and reconstruction approaches have been used to improve the quality of $^{23}$Na MRI images acquired with reduced scan times.[4] For example, statistical based methods and compressed sensing (CS) have been used to successfully denoise $^{23}$Na images of the brain[5] and leg.[6,7] However, these iterative techniques are computationally demanding, making their application time consuming and ill-suited to the clinical environment. Furthermore, the reduction in scan times attainable with these techniques is limited. Machine learning (ML) is now increasingly used to denoise and suppress artefacts in medical images,[8] with superior image quality compared to CS and much shorter reconstruction times.[9] Several studies have shown that ML can successfully denoise highly corrupted $^{1}$H images by leveraging the availability of large amounts of $^{1}$H training data.[10–12]

However, a similar approach cannot be taken for $^{23}$Na MRI, as large volumes of high-quality $^{23}$Na images are not readily available. A few studies have attempted to increase the amount of $^{23}$Na training data by taking a patch-based approach.[13,14] For instance, Adlung et al. trained a U-Net[15] for denoising 4× undersampled 3D $^{23}$Na images using two-dimensional (2D) patches taken from the central slices of 38 image volumes (9500 2D image patch in total).[13] This model outputted high SNR $^{23}$Na images but resulted in a TSC error of approximately 3.9 mM.

Here, we propose an alternative approach to ML denoising of $^{23}$Na images, using large amounts of readily available, high-quality $^{1}$H data to train a convolutional neural network (CNN). Although training and inference data often have the same distribution, there is evidence that different image characteristics in the training and inference data has little impact on network performance.[16,17] Furthermore, networks either pretrained or solely trained on natural images have been successfully used to reconstruct $^{1}$H images of the head and the heart.[18,19] Thus, we hypothesized that a network trained on $^{1}$H data could be used to successfully denoise prospective $^{23}$Na images of the calf.

We have previously demonstrated that a CS reconstruction of 2D $^{23}$Na data with 50 averages maintains SNR and TSC accuracy compared to reference images acquired with 150 averages.[7] In this study, we want to investigate if ML can achieve superior denoising compared to CS. We aimed to: i) use high quality $^{1}$H data to simulate paired high and low SNR $^{23}$Na images, for a range of signal averages, ii) train denoising CNNs using the synthetic paired datasets, and iii) test the denoising CNNs in prospectively acquired low-average $^{23}$Na images, by assessing image quality and accuracy of TSC quantification.

## 2  Materials and methods

### 2.1  Synthetic paired data
The open-source $^1$H fastMRI knee dataset[20] was used to create synthetic training data. To best mimic prospective $^{23}$Na calf images (i.e. low signal from fat and bone, high signal from blood), a subset of 1893 2D T$_2$-weighted axial slices acquired with a turbo spin echo pulse sequence with fat saturation were used.

Ground truth images with similar properties to prospective $^{23}$Na images were generated by first creating a square matrix using reflection padding and then downsampling to a matrix of 80×80 (Fig. 1A). A circle of high signal intensity, which mimicked the sodium reference phantoms used in prospective imaging, were added to $^1$H images. This was done to ensure the range of pixel intensities were comparable to those in prospective $^{23}$Na images following normalization. The diameter of the circle was 12 pixels, and the pixel intensity was 7.5× the mean image pixel intensity. This scaling was determined from prospective $^{23}$Na images by comparing the mean pixel intensity in the 100 mM phantom to the mean pixel intensity across the whole image. The added circle was blurred using a Gaussian filter ($\sigma$ = 0.6) prior to superimposition in a random corner of the image (between 9 and 15 pixels from the edges).

This image was converted to k-space using a fast Fourier transform (FFT), resampled using the spiral trajectory of the prospective data (spiral trajectory parameters: matrix = 80×80, field of view (FOV) = 180×180 mm$^2$, 20 regularly spaced spirals for complete filling of k-space, spiral readout duration = 7.6 ms) and reconstructed using a non-uniform FFT (NUFFT) with density compensation.[21] This image was used as the high SNR ground truth.

To create the synthetic paired low SNR data (Fig. 1A), multiple replicates of the resampled spiral k-space data were created, and Gaussian noise was added to each copy. The appropriate number of replicates (50, 40, 30, 20 or 10) were then averaged and reconstructed using a NUFFT with density compensation, as above. The level of noise was determined empirically, by comparing simulated $^1$H images reconstructed across a range of signal averages, with equivalent prospective $^{23}$Na images (Fig. 1B), and was defined with a standard deviation ($\sigma$) of 0.038× the maximum k-space value for each individual image.

### 2.2  Network architecture
A modified denoising CNN [22] architecture was used in this study (Fig. 1C). An initial 2D convolutional layer with 3×3 kernels and rectified linear unit (ReLU) activation was followed by eight residual blocks (Fig. 1C).[23,24] A final 2D convolutional layer (3×3 kernels), ReLU activation and batch normalization was followed by a 1×1 convolution with one filter to combine features into a single channel. A global residual addition was then used to output the denoised image. The number of residual blocks was optimized empirically.

### 2.3  Network training and validation
The synthetic dataset was normalized between 0 and 1 and randomly split between training (55%), validation (20%) and testing (25%). The denoising CNNs were implemented and trained using TensorFlow[25] with a joint loss consisting of the Jensen Shannon distance[26] between the residuals and the structural similarity index (SSIM) between the images, weighted

0.3:1.0 respectively. The weighting between the two loss components was determined empirically.

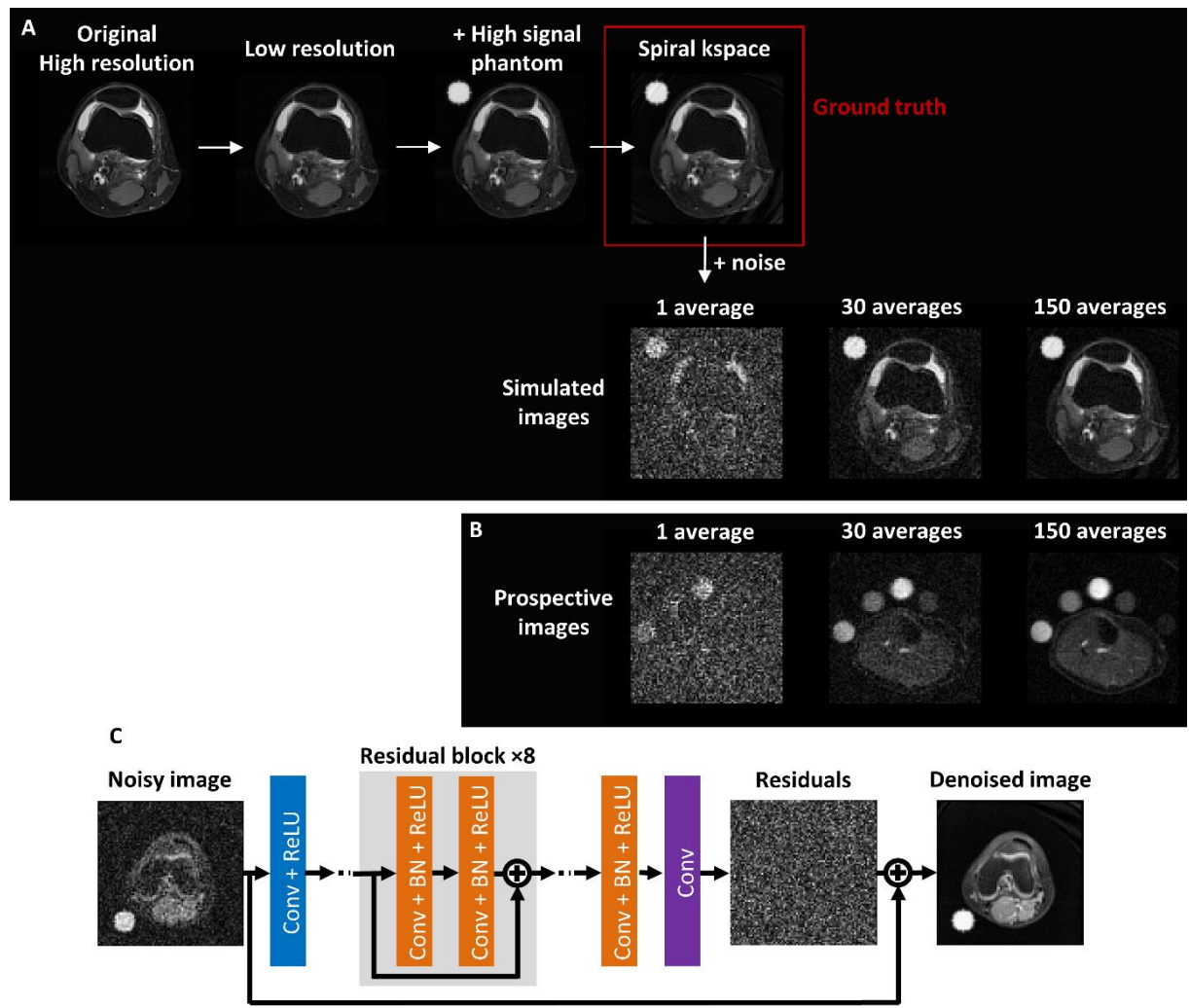

**Fig. 1.** A) Preparation of synthetic paired data using the open-source $^1$H fastMRI knee dataset.[20] A downsampled image with a simulated high-signal $^{23}$Na reference phantom was resampled with a spiral k-space trajectory and reconstructed to give the ground truth input to the denoising convolutional neural network (CNN). Gaussian noise was added to replicates of the low-resolution, resampled spiral k-space, resulting in comparable image quality to prospective data (B). A range of number of replicates were averaged and reconstructed to give the noisy input to the CNN. C) Architecture of denoising CNN. Conv: 2D convolution; BN: batch normalization; ReLU: rectified linear unit.

Adam[27] was used as the optimizer with a learning rate of $10^{-4}$. Training continued for 100 epochs, with a batch size of 16. Performance of the networks was assessed on the $^1$H test dataset using mean squared error (MSE), SSIM and peak SNR (PSNR) metrics. The model was trained independently for different noise levels, corresponding to 50, 40, 30, 20, and 10 averages.

## 2.4 Prospective imaging

$^{23}$Na images of the calf were acquired in 10 healthy volunteers, as previously described,[7] with approval from the local research ethics committee (Ref. 15/0041). Images were acquired using a 2D uniform-density center-out spiral trajectory and half-pulse excitation with the following parameters: FOV = 180×180 mm$^2$, pixel size = 2.25×2.25 mm$^2$, slice thickness = 30 mm, echo time = 0.23 ms, repetition time = 100 ms, spiral readout duration = 7.6 ms, 20 regularly spaced spirals for complete filling of k-space, and 150 averages. Reference calibration phantoms with known sodium concentrations (10, 25, 50, 75 and 100 mM) were positioned around the leg, within the imaging FOV.

The trained networks were applied to prospective $^{23}$Na data which were retrospectively reconstructed with fewer averages (50, 40, 30, 20 and 10) using a NUFFT and density compensation and normalized between 0 and 1. These data were also reconstructed with an optimized CS reconstruction[7] for comparison. All image reconstructions were performed using Python (version 3.8.10) and TensorFlow MRI (version 0.22.0).[28]

## 2.5 Prospective $^{23}$Na Image quality assessment

Quantitative image quality metrics were measured to assess prospective $^{23}$Na image quality. Metrics included estimated SNR and edge sharpness. Estimated SNR was calculated using the mean signal intensity in a region of interest (ROI) in the muscle and standard deviation of the noise from a background ROI. Edge sharpness was calculated by measuring the maximum gradient of pixel intensities across the 100 mM phantom border on TSC maps, as previously described.[7] These metrics were both calculated in Python.

Subjective image quality was independently assessed by two observers (MR and SBW). For a given volunteer and number of averages, the observer was presented with four images (reference, NUFFT, CS and CNN-denoised) and asked to rank them from best (1) to worst (4) based on overall image quality. Note, competition ranking was used, allowing images to be ranked equally, and images were fully anonymized and presented in a random order.

## 2.6 TSC quantification

To generate TSC maps, a linear calibration was performed using mean values from ROIs drawn in the center of sodium reference phantoms. TSC values, reported as apparent TSC (aTSC)[29] were then taken as the mean value within ROIs drawn in the muscle and skin using ITK-SNAP (version 3.8.0).[7,30]

## 2.7 Statistical analysis

All statistical analysis was performed using RStudio (Version 2022.7.2.576, RStudio, PBC, Boston, MA, USA). For continuous data, distributions were tested for normality using Shapiro-Wilk tests and as most data were not normally distributed, non-parametric tests were used. Metrics calculated on the $^1$H test data were compared using pairwise paired Wilcoxon signed-rank tests. For prospective data, estimated SNR and edge sharpness measurements were compared using a Friedman test, and significant results were followed with post hoc pairwise paired Wilcoxon signed-rank tests with Bonferroni correction to determine significant differences. Bland-Altman analyses of the muscle and skin aTSC values were performed between reference images and NUFFT, CS and CNN images. aTSC mean biases and limits of agreement were reported. Biases were tested for statistical significance using a one-sample t-test. Additionally, Pearson correlation coefficients were calculated. Subjective image quality rankings were compared using a pairwise paired t-test. All analyses were repeated across all investigated number of averages. Results are considered statistically significant for p<0.05.

## 3 Results

### 3.1 Network testing

The CNNs were successfully trained on $^1$H data over a period of 9–10 hours each and applied to $^1$H test data. Image metrics comparing low SNR input images and denoised output images with simulated ground truth images are shown in Table 1. The denoised images had significantly lower MSE (p<0.001) and higher SSIM and PSNR (p<0.001 for both) compared to input images across all investigated number of averages.

**Table 1**

Image quality metrics from synthetic test data set

| Number of averages | Image | MSE [×10$^{-3}$] | SSIM | PSNR [dB] |
|---|---|---|---|---|
| 50 | Input | 1.3 ± 0.1 | 0.77 ± 0.05 | 29.0 ± 0.5 |
|  | Denoised | *0.4 ± 0.1 | *0.92 ± 0.03 | *33.7 ± 1.0 |
| 40 | Input | 1.6 ± 0.2 | 0.73 ± 0.05 | 28.0 ± 0.5 |
|  | Denoised | *0.5 ± 0.1 | *0.91 ± 0.03 | *33.1 ± 1.0 |
| 30 | Input | 2.1 ± 0.2 | 0.68 ± 0.06 | 26.9 ± 0.5 |
|  | Denoised | *0.6 ± 0.1 | *0.90 ± 0.03 | *32.6 ± 0.9 |
| 20 | Input | 3.0 ± 0.4 | 0.61 ± 0.06 | 25.2 ± 0.5 |
|  | Denoised | *0.7 ± 0.2 | *0.88 ± 0.04 | *31.6 ± 1.0 |
| 10 | Input | 5.7 ± 0.7 | 0.48 ± 0.07 | 22.5 ± 0.6 |
|  | Denoised | *0.9 ± 0.2 | *0.84 ± 0.05 | *30.3 ± 1.1 |

Results are presented as median ± interquartile range. MSE: mean squared error; SSIM: structural similarity index; PSNR: peak signal-to-noise ratio.
* Indicates a statistically significant difference between input and denoised images (p<0.05).

### 3.2 Prospective $^{23}$Na data

Trained CNNs were successfully used to denoise prospective $^{23}$Na images of the calf reconstructed with different numbers of averages (Fig. 2). Application of the CNNs to the $^{23}$Na images took approximately 0.1 s, in addition to 0.2 s for the initial image reconstruction. The data was also reconstructed using CS, with a processing time of approximately 32.8 s.

#### 3.2.1 Image quality assessment

Quantitative image quality metrics are presented in Table 2. For 20-50 averages, estimated SNR was significantly higher in CNN-denoised images compared to reference (p≤0.023), CS (p=0.012) and NUFFT images (p=0.012). However, for 10 averages there was no significant difference between CNN-denoised and reference images (p=0.29). It should be noted that estimated SNR was also significantly higher for CS compared to NUFFT images (p=0.012), although it was generally lower compared to reference images (p≤ 0.035, except for 50 averages).

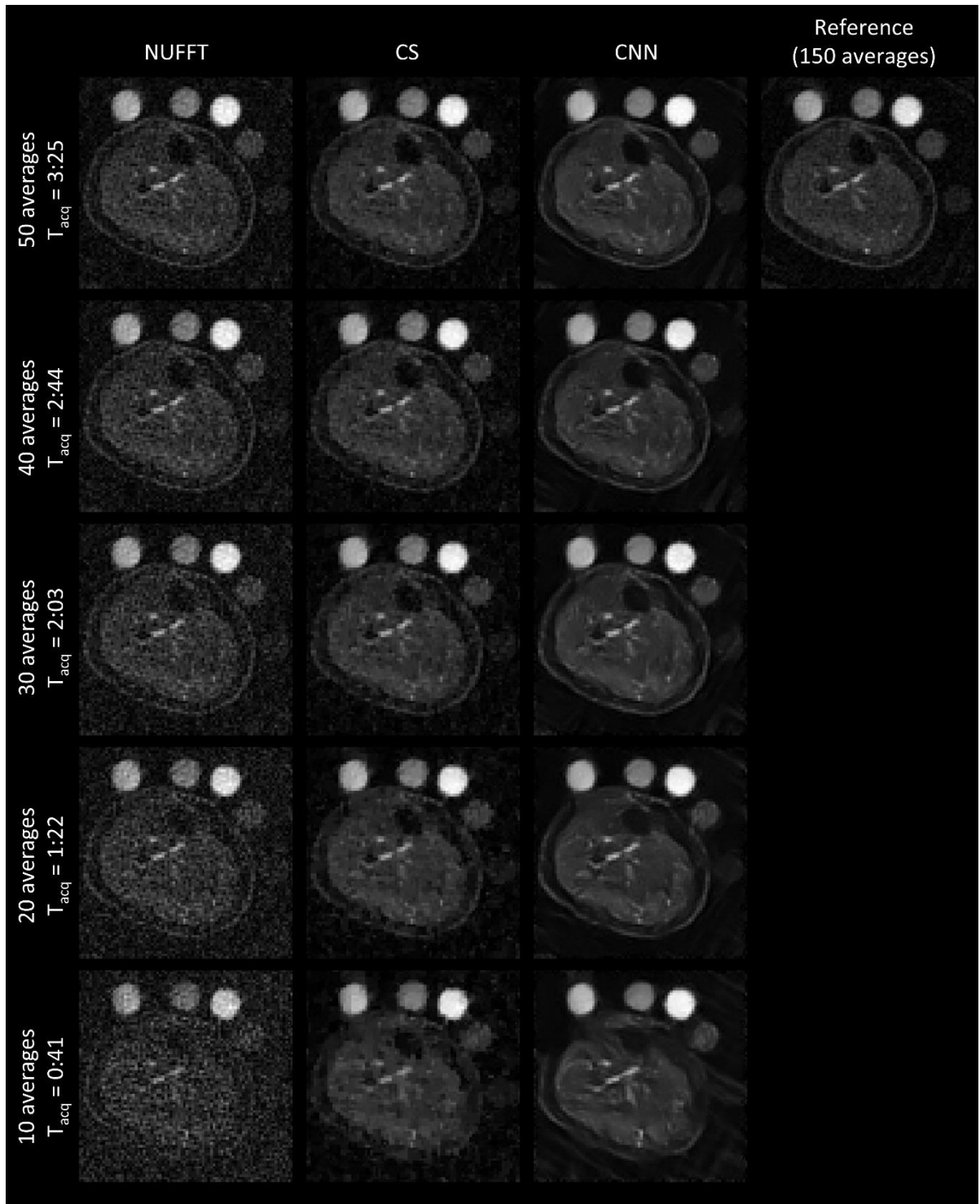

**Fig. 2.** Example of prospective $^{23}$Na in-vivo non-uniform fast Fourier transform (NUFFT) reference image (150 averages) and images retrospectively reconstructed with 50, 40, 30, 20 and 10 averages using NUFFT, compressed sensing (CS), and NUFFT with denoising convolution neural network (CNN) applied. Equivalent acquisition times (Tacq, min:s) are provided for each of the investigated number of averages.

Edge sharpness was similar for reference, NUFFT, CS and CNN-denoised images. Of the comparisons made, only two were statistically significant: with 30 averages, edge sharpness of CS images was significantly lower than NUFFT images (p=0.023), and with 50 averages, edge sharpness of CNN-denoised images was significantly higher than CS images (p=0.012).

**Table 2**

Prospective $^{23}$Na data quantitative image quality metrics

| Number of averages | Image | Estimated SNR | Edge sharpness [mM mm$^{-1}$] |
|---|---|---|---|
| 150 | Reference | 6.7 ± 1.2 | 12.6 ± 0.7 |
| 50 | NUFFT | *4.4 ± 0.5 | 12.6 ± 0.5 |
|  | CS | #6.8 ± 1.0 | 12.5 ± 0.6 |
|  | CNN | *#+24.0 ± 5.3 | +12.8 ± 0.5 |
| 40 | NUFFT | *4.1 ± 0.5 | 12.6 ± 0.4 |
|  | CS | *#5.8 ± 1.1 | 12.5 ± 0.4 |
|  | CNN | *#+20.4 ± 4.4 | 12.7 ± 0.5 |
| 30 | NUFFT | *3.6 ± 0.5 | 12.5 ± 0.9 |
|  | CS | *#5.9 ± 0.9 | #12.4 ± 0.8 |
|  | CNN | *#+21.2 ± 10.6 | 12.6 ± 0.5 |
| 20 | NUFFT | *2.9 ± 0.4 | 12.4 ± 1.3 |
|  | CS | *#5.5 ± 0.9 | 12.4 ± 1.0 |
|  | CNN | *#+22.6 ± 7.8 | 12.4 ± 0.5 |
| 10 | NUFFT | *2.2 ± 0.4 | 12.7 ± 2.8 |
|  | CS | *#5.5 ± 1.3 | 12.1 ± 1.2 |
|  | CNN | #+9.4 ± 4.3 | 12.4 ± 0.5 |

Results are presented as median ± interquartile range. SNR: signal-to-noise ratio; NUFFT: non-uniform fast Fourier transform; CS: compressed sensing; CNN: convolutional neural network.

\* Indicates a statistically significant difference (p<0.05) compared to Reference.
# Indicates a statistically significant difference (p<0.05) compared to the NUFFT.
+ Indicates a statistically significant difference (p<0.05) compared to the CS.

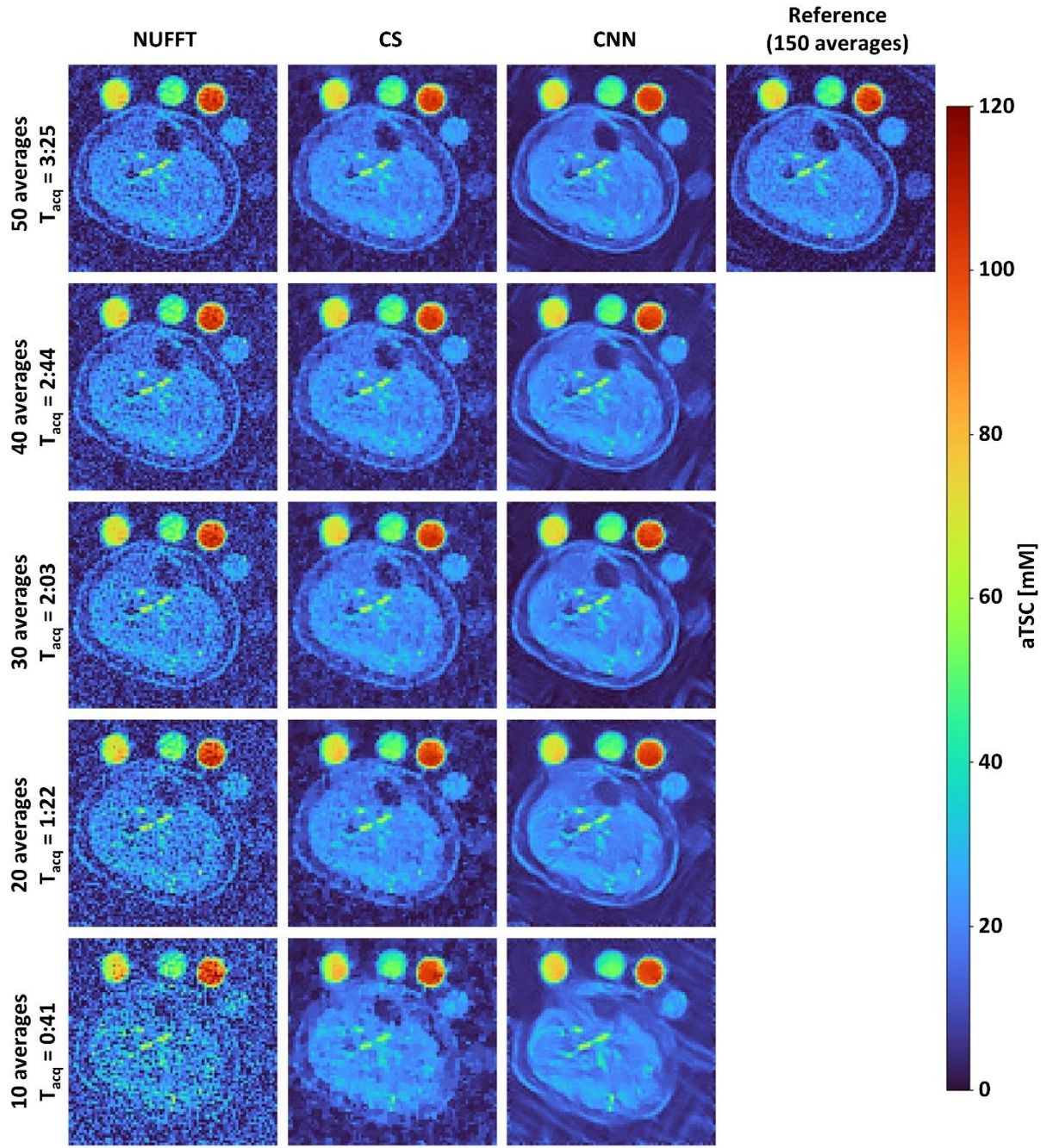

**Fig. 3.** Example of prospective $^{23}$Na in-vivo apparent tissue sodium concentration (aTSC) maps generated from a non-uniform fast Fourier transform (NUFFT) reference image (150 averages) and images retrospectively reconstructed with 50, 40, 30, 20 and 10 averages using NUFFT, compressed sensing (CS), and NUFFT with denoising convolution neural network (CNN) applied (corresponding to images presented in Fig. 2). Equivalent acquisition times (Tacq, min:s) are provided for each of the investigated number of averages.

Subjective image ranking results are presented in Table 3. The CNN-denoised images were ranked equally best with reference images (p=1.0) and significantly better than CS (p≤0.031) and NUFFT images (p≤0.015), for all investigated numbers of averages. CS images ranked significantly better than NUFFT images for 10 (p=0.006) and 20 (p=0.003) averages, but there was no significant difference between them for 50 (p=0.76), 40 (p=1.0) and 30 averages (p=0.36).

**Table 3**
Prospective $^{23}$Na data subjective image quality ranking

| Number of averages | Image | Image quality ranking |
|---|---|---|
| 50 | Reference | 1.7 ± 0.7 |
|  | NUFFT | *3.4 ± 1 |
|  | CS | *2.8 ± 0.8 |
|  | CNN | #+1.6 ± 0.9 |
| 40 | Reference | 1.7 ± 0.6 |
|  | NUFFT | *3.1 ± 1 |
|  | CS | *3 ± 0.7 |
|  | CNN | #+1.5 ± 0.8 |
| 30 | Reference | 1.6 ± 0.5 |
|  | NUFFT | *3.4 ± 0.8 |
|  | CS | *3 ± 0.5 |
|  | CNN | #+1.4 ± 0.7 |
| 20 | Reference | 1.6 ± 0.5 |
|  | NUFFT | *3.6 ± 0.6 |
|  | CS | *#2.8 ± 0.4 |
|  | CNN | #+1.4 ± 0.6 |
| 10 | Reference | 1.4 ± 0.5 |
|  | NUFFT | *3.1 ± 1 |
|  | CS | *#2.5 ± 0.5 |
|  | CNN | #+1.6 ± 0.6 |

Results are presented as mean ± standard deviation. Images are ranked from best (1) to worst (4). NUFFT: non-uniform fast Fourier transform; CS: compressed sensing; CNN: convolutional neural network.
* Indicates a statistically significant difference (p<0.05) compared to Reference.
# Indicates a statistically significant difference (p<0.05) compared to the NUFFT.
+ Indicates a statistically significant difference (p<0.05) compared to the CS.

### 3.2.2 aTSC quantification

The aTSC maps for all signal averages are presented in Fig. 3, and aTSC values and statistical results are presented in Tables 4 and 5.

Muscle aTSC values measured from both CS and CNN-denoised images were significantly higher (p≤0.006) than reference values, for all investigated numbers of averages. However, the observed bias was small (<1.4 mM and < 1.5 mM for CS and CNN-denoised images, respectively), and decreased with increasing number of averages (Table 4). No significant differences (p>0.05) were observed between muscle aTSC measured from NUFFT and reference images (bias <0.3 mM).

**Table 4**

Prospective muscle aTSC values and correlation and Bland-Altman analyses

| Number of averages | Image type | Muscle aTSC [mM] | Pearson correlation | | Bland-Altman | |
|---|---|---|---|---|---|---|
| | | | r | p | Bias (LOA) | p |
| 150 | Reference | 21.5 ± 3.0 | | | | |
| 50 | NUFFT | 21.6 ± 2.6 | 0.991 | <0.001 | 0.15 (-0.37-0.67) | 0.115 |
| | CS | 22.0 ± 2.7 | 0.986 | <0.001 | 0.67 (-0.02-1.36) | *<0.001 |
| | CNN | 22.0 ± 2.8 | 0.993 | <0.001 | 0.59 (0.06-1.12) | *<0.001 |
| 40 | NUFFT | 21.4 ± 1.7 | 0.963 | <0.001 | 0.12 (-0.90-1.14) | 0.484 |
| | CS | 22.0 ± 1.9 | 0.958 | <0.001 | 0.64 (-0.45-1.73) | *0.005 |
| | CNN | 22.1 ± 1.8 | 0.974 | <0.001 | 0.65 (-0.22-1.52) | *0.001 |
| 30 | NUFFT | 21.3 ± 1.7 | 0.967 | <0.001 | 0.05 (-1.06-1.16) | 0.784 |
| | CS | 22.1 ± 2.0 | 0.953 | <0.001 | 0.69 (-0.45-1.83) | *0.005 |
| | CNN | 21.8 ± 2.0 | 0.977 | <0.001 | 0.47 (-0.34-1.28) | *0.006 |
| 20 | NUFFT | 21.6 ± 1.8 | 0.931 | <0.001 | -0.23 (-1.64-1.17) | 0.328 |
| | CS | 22.5 ± 1.9 | 0.953 | <0.001 | 0.74 (-0.43-1.91) | *0.004 |
| | CNN | 23.0 ± 2.2 | 0.958 | <0.001 | 1.09 (-0.01-2.20) | *<0.001 |
| 10 | NUFFT | 21.4 ± 2.5 | 0.838 | 0.002 | -0.18 (-2.26-1.89) | 0.595 |
| | CS | 22.5 ± 2.6 | 0.883 | <0.001 | 1.31 (-0.46-3.08) | *0.001 |
| | CNN | 22.6 ± 2.4 | 0.834 | 0.003 | 1.44 (-0.66-3.53) | *0.002 |

Muscle aTSC values are reported as median ± interquartile range and Bland-Altman results are reported as bias (limits of agreement). aTSC: apparent tissue sodium concentration; LOA: limits of agreement; NUFFT: non-uniform fast Fourier transform; CS: compressed sensing; CNN: convolutional neural network.
* Indicates a statistically significant bias from 0 (p<0.05).

Skin aTSC values measured from CS images were significantly lower (p≤0.011) than reference values, for all investigated numbers of averages. Skin aTSC values measured from CNN-denoised maps were not significantly different (p>0.05) from reference values, except with 30 averages (p=0.021). As with muscle aTSC, the observed bias was small (<2.3 mM and <1.0 mM for CS and CNN-denoised images, respectively) and decreased with increasing number of averages (Table 5). No significant differences (p>0.05) were observed between skin aTSC measured from NUFFT and reference images (bias <0.2 mM).

In terms of balancing reduced scan time, image quality, and aTSC accuracy, 30 averages appears to be the best compromise. Correlation and Bland-Altman plots for images reconstructed with 30 averages are presented in Fig. 4.

**Table 5**
Prospective skin aTSC values and correlation and Bland-Altman analyses

| Number of averages | Image type | Skin aTSC [mM] | Pearson correlation | | Bland-Altman | |
|---|---|---|---|---|---|---|
| | | | r | p | Bias (LOA) | p |
| 150 | Reference | 18.8 ± 5.1 | | | | |
| 50 | NUFFT | 18.8 ± 5.7 | 0.981 | <0.001 | 0.15 (-1.06-1.37) | 0.449 |
| | CS | 17.5 ± 5.7 | 0.979 | <0.001 | -0.76 (-2.07-0.55) | *0.006 |
| | CNN | 18.3 ± 5.6 | 0.976 | <0.001 | -0.25 (-1.67-1.18) | 0.312 |
| 40 | NUFFT | 19.0 ± 5.5 | 0.970 | <0.001 | 0.05 (-1.48-1.59) | 0.829 |
| | CS | 17.7 ± 5.5 | 0.971 | <0.001 | -0.78 (-2.28-0.72) | *0.011 |
| | CNN | 18.4 ± 5.6 | 0.970 | <0.001 | -0.37 (-2.11-1.38) | 0.227 |
| 30 | NUFFT | 19.0 ± 5.5 | 0.979 | <0.001 | 0.04 (-1.27-1.34) | 0.866 |
| | CS | 17.4 ± 5.6 | 0.975 | <0.001 | -1.12 (-2.52-0.27) | *<0.001 |
| | CNN | 18.1 ± 6.1 | 0.973 | <0.001 | -0.84 (-2.7-1.02) | *0.021 |
| 20 | NUFFT | 18.7 ± 4.4 | 0.967 | <0.001 | -0.16 (-1.79-1.47) | 0.561 |
| | CS | 16.5 ± 5.0 | 0.954 | <0.001 | -1.69 (-3.66-0.28) | *<0.001 |
| | CNN | 17.9 ± 5.5 | 0.966 | <0.001 | -0.42 (-2.49-1.65) | 0.240 |
| 10 | NUFFT | 18.7 ± 4.5 | 0.918 | <0.001 | -0.14 (-2.65-2.37) | 0.738 |
| | CS | 16.4 ± 4.3 | 0.908 | <0.001 | -2.25 (-5.09-0.59) | *<0.001 |
| | CNN | 17.2 ± 6.6 | 0.879 | <0.001 | -0.98 (-4.96-2.99) | 0.159 |

Skin aTSC values are reported as median ± interquartile range and Bland-Altman results are reported as bias (limits of agreement). aTSC: apparent tissue sodium concentration; LOA: limits of agreement; NUFFT: non-uniform fast Fourier transform; CS: compressed sensing; CNN: convolutional neural network.
* Indicates a statistically significant bias from 0 (p<0.05).

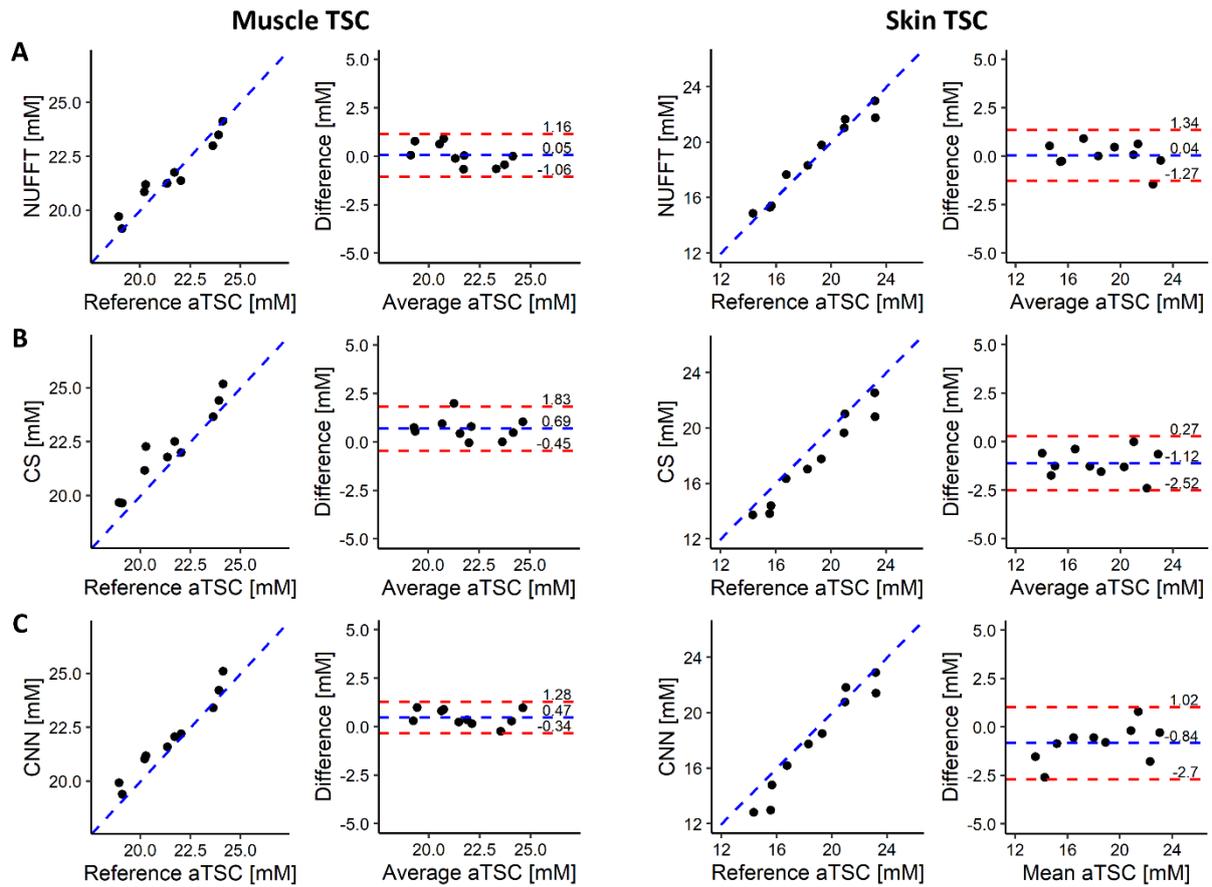

**Fig. 4.** Correlation and Bland-Altman plots comparing muscle and skin aTSC values determined from prospective $^{23}$Na images retrospectively reconstructed with 30 averages using a non-uniform fast Fourier transform (NUFFT) (A), compressed sensing (CS) (B) and NUFFT with denoising convolution neural network (CNN) (C) to reference values determined from 150 average data.

## 4  Discussion

In this study, we trained a CNN to denoise $^{23}$Na images of the calf. The main findings of the study are: i) it is possible to train a denoising CNN on $^1$H data and successfully apply the resultant network to prospective $^{23}$Na images; ii) CNN-denoised images have better image quality than CS images and can be reconstructed 100× faster; iii) muscle and skin aTSC values determined from CNN-denoised images are equivalent to those from CS images and differ from reference values by less than 0.9 mM.

### *4.1  Network and training*
To our knowledge, this is the first time a CNN trained on $^1$H data has been used to denoise $^{23}$Na images. Previous ML approaches for $^{23}$Na MRI denoising used patch-based methods that make the best use of the small amounts of training data.[13,14] However, intra-patient correlation between patches may limit the amount of 'new information' the model is exposed to during training, potentially affecting robustness. Instead, we used $^1$H data to simulate $^{23}$Na MRI-like images for training, leveraging large publicly available $^1$H MRI repositories suitable for this purpose.[31] Specifically, we used the fastMRI knee dataset,[20] resulting in a training dataset up to 40× larger than ML models previously applied to $^{23}$Na MRI, without the need for patching.[6,13]

In this study, we used an architecture based on the denoising CNN first proposed by Zhang et al., which has been shown to effectively denoise natural images.[22] Importantly, variations of the network have previously been used to successfully denoise $^1$H MRI images of the brain.[10,32] Features of our network include a global residual and internal residual blocks, which enable deeper networks to be trained without diminished accuracy.[23] We also utilized a regularized loss function that included SSIM and the Jensen Shannon distance. The latter is often used in generative adversarial networks[33] to ensure similarity between input and output distributions. In our study, the addition of the Jensen Shannon distance helped to prevent removal of true signal in hyperintense regions of the image, particularly the high concentration sodium phantoms. On the synthetic data, our model effectively removed noise while maintaining edge sharpness. However, the more important test was to evaluate the effectiveness of the model in denoising prospective $^{23}$Na MRI data.

### 4.2  Prospective $^{23}$Na data and comparison with CS

We demonstrated that our CNN trained on $^1$H data was able to successfully denoise prospective $^{23}$Na images of the calf with a reduced number of averages. Images reconstructed with ≥ 30 averages showed good image quality and accurate aTSC quantification. This proves that for our application, training with $^1$H data does not result in reduced generalizability when denoising $^{23}$Na images. This is in line with recent work that has demonstrated that training data for denoising/artefact suppression can superficially have very different image characteristics.[16,17,19]

Interestingly, estimated SNR was higher in the CNN-denoised images than reference $^{23}$Na images acquired with 150 averages. This was probably due to the higher SNR of the $^1$H ground truth training data, compared to $^{23}$Na reference images. Importantly, edge sharpness was equivalent in CNN-denoised, NUFFT and reference images, suggesting no blurring was introduced by the denoising CNN, a common problem for other denoising approaches.

We showed that muscle and skin aTSC from CNN-denoised images were in very good agreement with the reference data with biases of <0.9 mM. This is less than previously reported for patch-based CNN-denoising of $^{23}$Na images and supports our underlying approach. Our CNNs were able to successfully denoise images with as little as 30 averages without blurring or artefacts. Below this number of averages, CNN denoising resulted in visible blurring of small/thin structures, the introduction of artefacts and errors in aTSC values. Nevertheless, 30 averages represents a 5× speed up compared to the reference scan enabling an image to be acquired in 2 minutes (compared to 10 minutes conventionally).

Previously, we have shown that CS can also successfully denoise $^{23}$Na images with reduced signal averages. However, the ML approach performed better than CS, in terms of effective denoising and processing time. This is reflected in the estimated SNR measurements, as well as subjective image quality ranking. The bias in muscle aTSC values was similar for CS and CNN-denoised images. It should be noted that the bias in skin aTSC was slightly worse for CS than ML, but the limits of agreement for the CNN-denoised skin aTSC values were slightly wider than for CS. A significant advantage of ML-denoising is the fast processing time, which in this study was just 0.3 s (including the NUFFT image reconstruction), which is 100× faster than the CS reconstruction.

### 4.3  Limitations and future work

In this study, the network was trained on magnitude data; this was due to the DICOM format of the $^1$H data used for training. The fastMRI knee dataset does contain some raw k-space data, but these are solely for images acquired in the coronal plane. Since our prospective $^{23}$Na images were acquired transversely, we decided to use the transverse DICOM data for training. Alternatively, complex data may be used, as previously done for accelerated MRI reconstruction[34] and deep artifact suppression,[35] which may improve the performance of the network further.

Furthermore, our denoising CNN was a postprocessing tool, applied to $^{23}$Na images reconstructed using a NUFFT. Reconstruction models such as a variational network (VarNet)[36,37] have been shown to effectively reconstruct $^1$H MRI images from input multi-coil or single-coil data.[38] Using such networks may enable further acceleration of $^{23}$Na MRI scans, while also maintaining data consistency. The method by which the $^{23}$Na acquisitions are accelerated may also be investigated further; instead of reducing the number of averages (with a fully sampled k-space), the k-space trajectory may alternatively be undersampled, as is typically done with $^1$H or 3D $^{23}$Na MRI.

# 5 Conclusion

We have shown that CNNs can be trained on high quality $^1$H data and subsequently used to denoise $^{23}$Na images of the calf. With improved performance compared to CS, the denoising-CNN enables $^{23}$Na images to be acquired within 2 minutes, with a bias in aTSC of no more than ± 0.9 mM compared to images acquired over 10 minutes. This reduction in scan time will make $^{23}$Na MRI acquisitions more suitable for use clinically, in large-cohort and longitudinal research studies.

**Funding:** RRB is funded by The Michael J. Fox Foundation for Parkinson's Research (grant MJFF-021438). JAS is funded by the UK Research and Innovation (UKRI) Future Leaders Fellowship (grant MR/S032290/1).